\documentclass[sigconf]{acmart}

\copyrightyear{2026}
\acmYear{2026}
\setcopyright{cc}
\setcctype{by}
\acmConference[MM '26]{Proceedings of the 34th ACM International Conference on Multimedia}{November 10--14, 2026}{Rio de Janeiro, Brazil}
\acmBooktitle{Proceedings of the 34th ACM International Conference on Multimedia (MM '26), November 10--14, 2026, Rio de Janeiro, Brazil}
\acmDOI{10.1145/3767308.3834750}
\acmISBN{979-8-4007-2213-4/2026/11}

\usepackage{amssymb}      
\usepackage{booktabs}
\usepackage{url}          
\usepackage{balance}      
\usepackage{placeins}     

\DeclareUrlCommand\codepath{\urlstyle{tt}}

\usepackage{xcolor}
\newcommand{\optfig}[5][0.85\linewidth]{%
  \begin{figure}[t]
    \centering
    \IfFileExists{#2}{%
      \includegraphics[width=#1]{#2}%
    }{%
      \fbox{\parbox{#1}{\centering\vspace{0.6em}%
        \textcolor{red!70!black}{\textbf{[Missing figure: \texttt{#2}]}}\\[0.3em]%
        \footnotesize Drop the screenshot at this path and rebuild.\vspace{0.6em}}}%
    }
    \caption{#3}
    \Description{#4}
    \label{#5}
  \end{figure}
}
\newcommand{\optfigh}[5][0.85\linewidth]{%
  \begin{figure}[ht]
    \centering
    \IfFileExists{#2}{%
      \includegraphics[width=#1]{#2}%
    }{%
      \fbox{\parbox{#1}{\centering\vspace{0.6em}%
        \textcolor{red!70!black}{\textbf{[Missing figure: \texttt{#2}]}}\\[0.3em]%
        \footnotesize Drop the screenshot at this path and rebuild.\vspace{0.6em}}}%
    }
    \caption{#3}
    \Description{#4}
    \label{#5}
  \end{figure}
}
\newcommand{\optfigwide}[5][\linewidth]{%
  \begin{figure*}[t]
    \centering
    \IfFileExists{#2}{%
      \includegraphics[width=#1]{#2}%
    }{%
      \fbox{\parbox{0.95\linewidth}{\centering\vspace{0.6em}%
        \textcolor{red!70!black}{\textbf{[Missing figure: \texttt{#2}]}}\\[0.3em]%
        \footnotesize Drop the screenshot at this path and rebuild.\vspace{0.6em}}}%
    }
    \caption{#3}
    \Description{#4}
    \label{#5}
  \end{figure*}
}

\title{TimeCues Studio: A Workspace for Music Annotation and Algorithm Prototyping}

\author{Sapir Caduri}
\email{sapir.caduri@gmail.com}
\affiliation{%
  \institution{Bar-Ilan University}
  \city{Ramat Gan}
  \country{Israel}}

\author{Yoav Goldberg}
\email{yoav.goldberg@cs.biu.ac.il}
\affiliation{%
  \institution{Bar-Ilan University}
  \city{Ramat Gan}
  \country{Israel}}

\renewcommand{\shortauthors}{Sapir Caduri and Yoav Goldberg}

\begin{document}

\begin{abstract}
Multimedia applications require precise music annotation---labeled positions, segments, or loops---placed by hand or algorithmically. Machine-learning algorithms are scalable and effective but need annotated training data, scarce for many tasks. TimeCues Studio is an open-source workspace where algorithm-development teams annotate a music corpus, compare detection algorithms against those annotations, and prototype new ones. Unlike existing tools built for a single track at a time, TimeCues targets teams annotating whole collections, tightly integrated with algorithm development. Annotators place several marker types---each supporting ambiguity-aware labeling---on a grid-locked timeline that visualizes many music features, including separated audio stems. The same timeline drives an algorithm-comparison engine with bundled baselines, a Python sandbox for prototyping new models, and an ambiguity-aware evaluator that honors the structured fields. The same visualization suits solo annotators on music-sync projects. TimeCues is MIT-licensed and deploys via one Docker Compose command.
\end{abstract}



\maketitle

\optfigwide{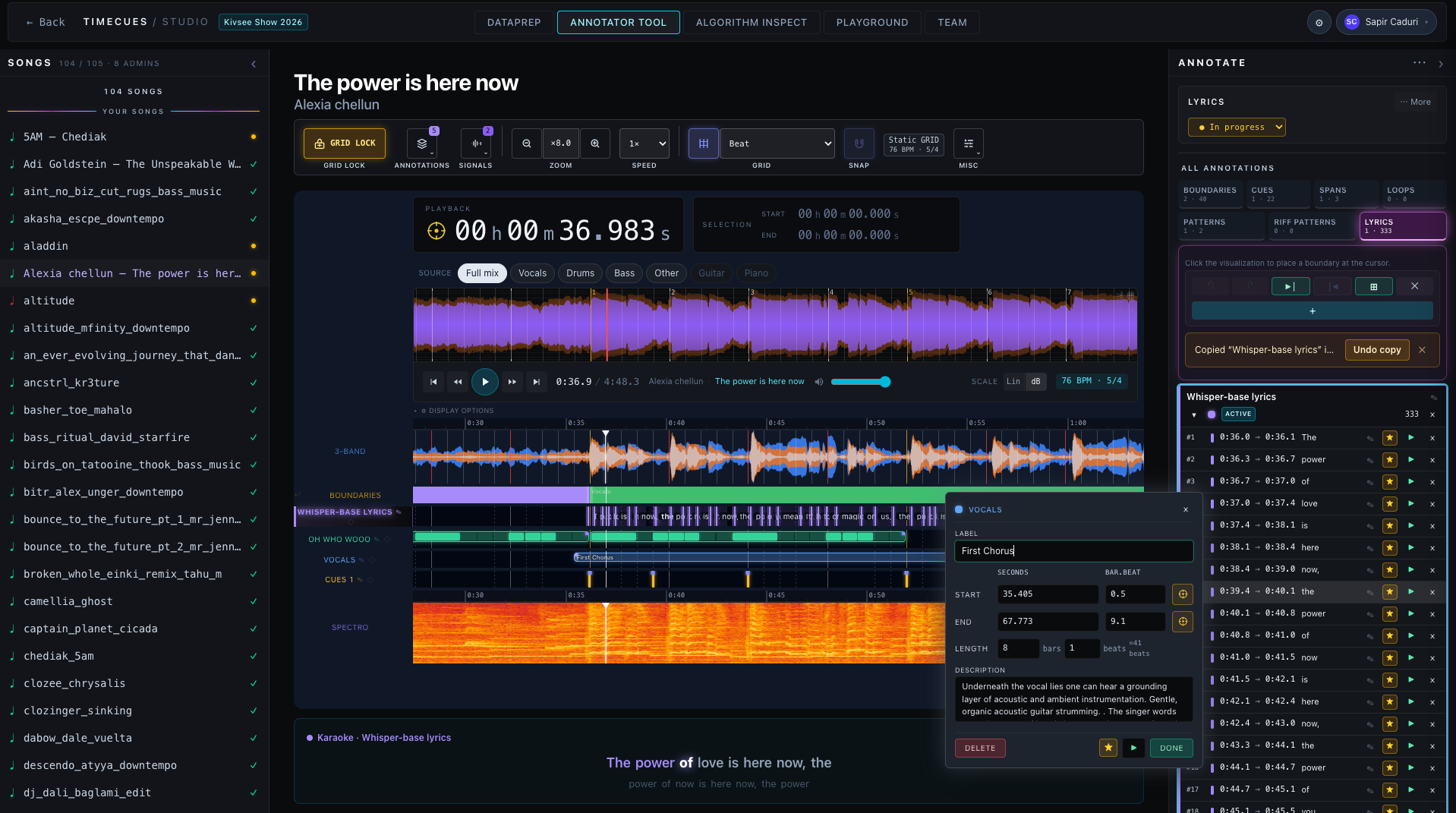}{%
  TimeCues Annotator Tool. \textit{Top:} layer pickers, zoom, beat-grid and snap controls (76\,BPM here). \textit{Left:} songs sidebar. \textit{Center:} audio visualization with annotation layers, a floating edit card, and an active-layer card with thumbnails. \textit{Right:} \textsc{Annotate} sidebar with marker tabs and marker list.}{Annotated screenshot of the TimeCues Annotator Tool: corpus sidebar (left), stacked beat-locked canvas (center) with an expanded layers (center), and the Annotate sidebar (right).}{fig:annotator}

\section{Introduction}
\label{sec:intro}

Automated music analysis is foundational to many multimedia applications and has been studied over the decades through a wide range of machine-learning models, supporting use cases that span automatic DJ mixing and track navigation to live multi-modal synchronization between music and visualization, such as stage lighting and visual effects~\cite{muller2015fundamentals, nieto2020audio}. These downstream uses of music synchronization demand very precise transitions: a lighting cue or video transition that misses the boundary is immediately visible and audible to the audience. Despite years of work on audio-driven show control---from real-time beat tracking and feature-extraction pipelines for stage lighting~\cite{goto2001audio, GangBLRHB11} to recent deep-learning attempts at end-to-end sound-to-DMX mapping~\cite{hils2023deep}---automatically synchronizing visuals to music remains an open problem. Solving it requires annotations at multiple granularities: coarse segment boundaries that mark where tension builds, releases, or shifts mood, and fine-grained markers that pin small sonic events---a snare hit, a transient, a vocal entry---to the minor visual accents synchronized with them.

Progress requires labeled corpora tuned to these timing demands and to a wide range of labeling tasks, plus an evaluation environment in which predictions can be inspected against those references. Existing annotation software is built for editing individual tracks~\cite{cannam2010sonic, boersma2001praat}; even when repurposed for machine learning, it lacks corpus management, team coordination, and built-in algorithm comparison, so each algorithmic evaluation requires a pipeline of separate tools. These tools also provide no built-in support for speeding up annotation with algorithm predictions that need only light manual correction, or for injecting one's own detection scripts. We present \textbf{TimeCues Studio}, an open-source web application that unifies corpus annotation, team coordination, and algorithm evaluation, analysis and comparison in a single workspace. We now describe its key principles and features.

\paragraph{Team collaboration.} For corpus creation and management, an admin assembles the track list (left panel of Figure~\ref{fig:annotator}), refines the suggested BPM, and re-aligns the beat grid~\cite{mcfee2015librosa}, then hands the prepared corpus to a team that annotates every track against that shared grid. A `Team dashboard' shows per-annotator progress, pairwise inter-annotator agreement, and access controls.

\paragraph{Flexible annotation markers.} To fit different music analysis tasks, TimeCues supports several marker types, letting users place exact segment boundaries, single time cues, overlapping time spans, and DJ-oriented loops and patterns directly on the metric grid. For exact details about the various annotation markers, see Section~\ref{sec:annotation}.
\paragraph{Ambiguity-aware annotations.} Annotation can be ambiguous: a section boundary can have several valid starting points~\cite{smith2011salami}, while points in between are not acceptable at all. In playlist generation, a transition might start at multiple points, all valid. In interactive audio looping, a producer might cue a section from either the start of a vocal pickup or the first heavy downbeat. An algorithm should be credited for hitting any candidate and penalized only when it misses all of them. Existing tools record a single timestamp per boundary and score it against a fixed tolerance window~\cite{raffel2014mireval, nieto2020audio}, or rely on frame-matching metrics~\cite{mcfee2015librosa}, which fail to handle discrete multi-candidate boundaries gracefully. 

In TimeCues Studio, we introduce an extended scheme that lets annotators mark all valid candidates for a single boundary, and our evaluator credits a match against any of them. The same multi-candidate logic extends to spans and cues through \emph{layers}: markers grouped in one layer are treated as semantically equivalent---say, three candidate vocal-entry cues (breath, consonant attack, vowel onset), or three competing phrasings of an instrumental fill---and the evaluator credits any prediction that hits a member. Like boundaries, each item carries a critical-versus-optional flag, so the dashboard can score the moments that matter most separately.

\paragraph{Algorithm development and evaluation.} \textbf{TimeCues provides a unified environment for manual annotation, semi\-automated annotation, and algorithmic evaluation}. In this single workspace, users can quickly prototype their own models via a built-in Python editor and compare them against an engine running diverse algorithmic baselines side by side on a shared timeline (Figure~\ref{fig:algo-inspect}), enabling rapid diagnostic feedback during hypothesis iteration. The engine exposes an extensibility layer for user-authored Python scripts, termed \emph{Custom Detectors}, which inject hypotheses onto the timeline. Beyond algorithmic evaluation, this mechanism serves as a semi-automated annotation assistant: scripts suggest candidate markers that annotators accept or reject in the annotation tool, accelerating curation.

\paragraph{Testing and scoring models.} The evaluation dashboard (Figure~\ref{fig:algo-inspect}) ships a suite of bundled detection algorithms from several families~\cite{nieto2016msaf, truong2020ruptures, kim2023allin1} and scores their predictions against any chosen reference. Alongside the metrics from \texttt{mir\_eval}~\cite{raffel2014mireval}---the de facto Python library for music-analysis evaluation---we add an \textit{ambiguity scheme} evaluator that credits matches against any candidate in a multi-candidate boundary or layer, weights optional alternatives by a tunable factor, and reports critical-only metrics. 

The same engine also computes \emph{AutoGuess} (Section~\ref{sec:autoguess}), a consensus that clusters several algorithms' predictions into high\-agree\-ment candidates---semi\-auto\-mated annotation suggestions that double as a tunable baseline.

\paragraph{Rich visualization.} TimeCues is built on the premise that visualization is central to both music annotation and algorithm analysis, and so supports a wide, flexible set of visual layers. The annotation and inspection workspaces share a single visual component (the center canvas of Figure~\ref{fig:annotator}): stacked, synchronized feature layers over one zoom-and-scroll timeline, a beat grid that markers can snap to so boundaries line up cleanly with the beat, and per-stem rendering so entries and exits can be seen on screen instead of being found by listening over and over.

In EDM, mashup, and DJ-style annotation tasks---a domain whose tempo and spectral structure (the distribution of energy across frequency bands) have been studied extensively for genre and subgenre characterization~\cite{caparrini2020automatic}---structural boundaries are almost always band-level events---structural changes that happen within a specific frequency band (bass, mids, or treble) rather than across the spectrum as a whole. A drop is bass suddenly appearing; a buildup is treble swelling; a breakdown is bass thinning. Locating these events on a standard monochrome amplitude waveform requires repeated listening, since the waveform collapses all frequencies into a single silhouette, while a spectrogram swings the other way, packing every frequency into dense per-pixel detail that is too noisy to scan at a glance. We therefore set the frequency-colored 3-band waveform as the default visualization (top of the center canvas in Figure~\ref{fig:annotator}, labeled \textsc{3-Band}): bass, mids, and treble are drawn in separate colors and stacked into one silhouette, so band-level events can be located by sight rather than by repeated listening. The 3-band view is a longstanding staple of commercial DJ software, adapted here from production tools~\cite{ni_traktor_waveform, pioneer_rekordbox_waveform, mixxx_waveforms}, but has not been adopted by open-source annotation tools, which still inherit the monochrome waveform or the dense spectrogram from generic audio editors~\cite{cannam2010sonic, boersma2001praat}.

\paragraph{Open and extensible.} TimeCues Studio is MIT-licensed and built for extension. A settings dashboard exposes configuration options such as default-value control and custom taxonomy definition; a single Docker Compose command brings up the full system. The container ships a small demo corpus of three CC0-licensed tracks\footnote{\emph{EDM At Midnight} (Play House); \emph{Pantheon} and \emph{Phonk Remix} (HoliznaCC0). All CC0~1.0 Universal, from the Free Music Archive (\url{https://freemusicarchive.org}).} so the tool can be tried end-to-end without supplying audio first.

\section{Related Work}
\label{sec:related}
\textbf{Annotation software.} The standard tools work one track at a time: Sonic Visualiser~\cite{cannam2010sonic} and Praat~\cite{boersma2001praat} have no notion of a corpus, a team, or an algorithm in the loop. Web tools widen the surface but still miss musical structure---Songle~\cite{goto2011songle} and Tony~\cite{mauch2015tony} target active listening and pitch, while audino~\cite{grover2020audino}, GECKO~\cite{levy2019gecko}, and BAT~\cite{melendez2017bat} target tags and transcripts. The closest, LabelBuddy~\cite{prokopiou2026labelbuddy}, frames annotation as verification over containerized back-ends; TimeCues takes that further with browser-authored custom detectors and a structure-aware scheme.

\textbf{Human-in-the-loop, evaluation \& consensus.} TimeCues's accept\slash reject workflow follows a long line of human-in-the-loop audio annotation systems~\cite{kim2018hitl,yamamoto2021arsa,pinto2025maracatu}. \texttt{mir\_eval}~\cite{raffel2014mireval} records a single timestamp per reference, yet SALAMI~\cite{smith2011salami} and McFee et al.~\cite{mcfee2017hierarchy} both show that choice is arbitrary; and when annotators disagree, the field aggregates labels~\cite{wang2017crowdsourcing} by methods from consensus clustering~\cite{ren2017consensus} (the basis of our AutoGuess) to IRT weighting~\cite{nakano2024irt}. TimeCues joins these threads, coupling scheme-aware scoring with live inter-annotator agreement (IAA).

\textbf{EDM \& show control.} EDM has drawn focused study---drops~\cite{yadati2014drops}, peak structure~\cite{solberg2019peak}, subgenre and tempo~\cite{caparrini2020automatic}, timbre~\cite{rocha2013segmentation}, and downbeats~\cite{hockman2012jungle}---while a growing set of downstream consumers, from beat\-tracked stage lighting~\cite{goto2001audio,GangBLRHB11} and sound\-to\-DMX~\cite{hils2023deep} to Songle Sync~\cite{kato2018songlesync} and ConcertCue~\cite{egozy2019concertcue}, need the authoring environment TimeCues provides.

\optfig[0.9\linewidth]{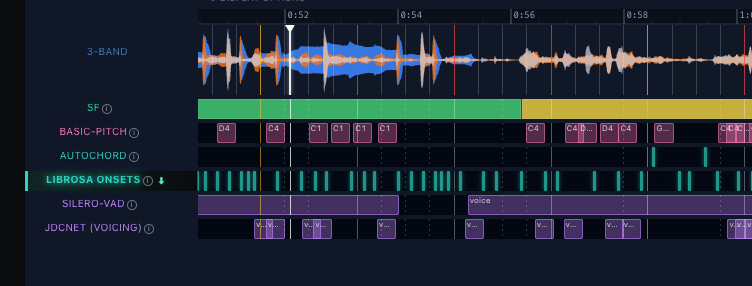}{%
  Algorithm Inspect: each detector gets its own row beneath the shared 3-band waveform}{Detector rows---Chroma Loops, Librosa Onsets, HPSS Percussive, and OLDA---stacked under the shared 3-band waveform.}{fig:algo-inspect}

\section{Annotation Lifecycle}
\label{sec:annotation}
TimeCues lets a team annotate a whole corpus in one place. An admin signs in, prepares the data, reviewers tag the songs on their own with per-track progress visible in the corpus sidebar (left panel of Figure~\ref{fig:annotator}), and a Team Dashboard surfaces team-level progress and how well annotators agree.

\subsection{Dataset Preparation}
An admin uploads a folder of audio files and sets a beat grid for each song. The importer accepts a mixed folder of audio and any existing metadata or annotations, previewing each decision before writing to disk. The grid runs in one of three modes: \emph{Static BPM} for a fixed tempo, with five estimators~\cite{mcfee2015librosa, bock2016madmom} suggesting values as chips; \emph{Dynamic} for tempos that drift, with an editable tempo curve; or \emph{Manual} for live or stitched takes, where each beat line can be placed by hand. A metronome plays so the grid can be checked by ear. Reviewers are then invited into the Annotator Tool (Figure~\ref{fig:annotator}).

The same page is where the dataset is managed: storage can be inspected and freed, and the full set of annotations, audio, stems, and algorithm outputs can be exported as JSON, Audacity~\cite{audacity}, Sonic Visualiser~\cite{cannam2010sonic}, JAMS~\cite{humphrey2014jams}, MIDI~\cite{midi_spec}, or REAPER~\cite{reaper}, and imported from JSON, Audacity, JAMS, and CSV.

\subsection{Annotator Tool}
\label{sec:annotator-tool}
The annotator opens a song and listens while the audio visualization scrolls across a timeline locked to the grid (center canvas of Figure~\ref{fig:annotator}). Annotation is organized into \textbf{layers}: a layer is a group of \textbf{markers} that share one type and meaning. Layers can be added as the task requires, with one tab per marker type at the top of the right panel in Figure~\ref{fig:annotator}: single points in time (\textbf{cues}), regions that do not overlap (\textbf{boundaries}), regions that can overlap (\textbf{spans}), or repeated regions (DJ-style \textbf{loops}\slash \textbf{patterns}).

The tool is built to cut work, save time, and reduce guesswork. Markers are placed with the mouse or keyboard (pressing \texttt{?} shows the full shortcut list). The annotator can zoom all stacked visualizations at once down to the sample, then drag-select a region and play it on repeat to isolate an event by ear. Labels and descriptions are editable; regions can be dragged, trimmed, or moved; markers snap to the grid to line up with the producer's beats in DAW-made music. The floating edit card overlaid on the center canvas in Figure~\ref{fig:annotator} holds the structured fields described in Section~\ref{sec:annotation-scheme}. Every change auto-saves, with stepwise undo and redo.

Different analysis and annotation tasks need different signals. The default view is the \textbf{3-band waveform}, which colors bass, mids, and treble separately. To our knowledge no other open-source annotator does this, and it makes events like an EDM kick pattern or a section change easy to spot. Beside it, the annotator can stack additional signals from the standard music-analysis feature set~\cite{muller2015fundamentals, mcfee2015librosa} (such as spectrogram, MFCC, chroma, and others), toggled from the \textsc{Signals} picker. Any of these signals can also be rendered for a single Demucs stem---Demucs~\cite{defossez2019demucs}  is a neural source-separation model that splits a mix into vocals, drums, bass, and \emph{other}, so the annotator can, for example, view the chroma of the vocals alone or the spectrogram of just the drum stem. Demucs runs built-in and the stem is chosen from a per-stem source selector at the top of the center canvas (\textsc{Source:} Full mix \slash Vocals \slash Drums \slash Bass \slash Other).

Annotators do not need to start from a blank canvas. A \emph{Custom Detector} (Section~\ref{sec:custom-detectors}) pre-fills a layer with algorithm suggestions, turning the task into accept\slash reject. \emph{AutoGuess} (Section~\ref{sec:autoguess}) takes the consensus across a chosen set of algorithms and surfaces the clusters as review cards on the timeline. Markers from either source can be copied into the manual layer, where they become part of the gold annotation. This speeds up labeling while keeping the manual layer a fully human-verified gold reference: because every algorithm carries some error or tolerance window, each copied marker is still vetted and adjusted by the annotator rather than trusted as-is. Each layer has a status (\emph{Not started}\slash \emph{In progress}\slash \emph{Reviewed}) and a per-marker stopwatch, so time spent is recorded alongside the label.

\paragraph{Extended-Annotation.}
\label{sec:annotation-scheme}
Each marker carries two fields beyond its label, both used by the evaluator (Section~\ref{sec:eval}). \textbf{Multi-candidate} captures events with several valid timestamps---a sung phrase may legitimately start at the breath, the consonant, or the vowel: cues and boundaries record every valid moment, and a prediction counts as a hit if it matches any of them. Region markers (spans, loops, and patterns) cover a whole region, not just a moment---an entire alternative loop or pattern can be equally valid---so they express this equivalence by grouping candidates in a \emph{layer} rather than as timestamps on one marker. \textbf{Criticality} (\emph{critical}\slash \emph{optional}) flags events whose timing tolerance differs---a drop must land exactly, a side fill can drift. Every marker is critical by default; the annotator clears the star on a thumbnail in the active layer card (bottom of the center canvas in Figure~\ref{fig:annotator}) to downgrade it to optional, and the metrics in Section~\ref{sec:eval} score the critical ones separately.

\section{Model Development Lifecycle}
\label{sec:eval}

Once a corpus has been annotated, the same workspace becomes a place to develop and compare detection models against those annotations---same audio, grid, feature cache, and canvas; only the dashboard changes. A researcher integrates a new model as a Custom Detector and applies it to some or all tracks.

Algorithm Inspect (Figure~\ref{fig:algo-inspect}) runs a set of bundled baselines that span the major families of music structure analysis (MSA): early novelty-curve segmentation~\cite{foote2000automatic}, feature-based MSAF pipelines~\cite{nieto2016msaf,nieto2013cnmf,vaz2016convex,mcfee2014learning,mcfee2014olda} and tensor decompositions~\cite{marmoret2020ntd}, change-point detection~\cite{truong2020ruptures,truong2018ruptures}, and deep models on demixed audio~\cite{kim2023allin1,hao2026songformerscalingmusicstructure}. It puts a representative of each on one canvas. The Custom Detector hook (Section~\ref{sec:custom-detectors}) welcomes new entrants.

Detection is not limited to boundaries. Beat and onset cues come from librosa and madmom~\cite{mcfee2015librosa, bock2016madmom}, and an optional \emph{experimental-models} profile adds opt-in detectors for the remaining marker types, each a separate sidecar gated behind a per-family setting: downbeat and meter tracking (BeatNet~\cite{heydari2021beatnet}), Krumhansl--Schmuckler key estimation~\cite{krumhansl1990cognitive}, chord recognition, and polyphonic note transcription (Basic Pitch~\cite{bittner2022basicpitch}) for cues; voice-activity (Silero-VAD~\cite{silerovad}) and singing-voice detection (JDC-Net~\cite{kum2019jdc}), AudioSet event tagging (PANNs~\cite{kong2020panns}), and percussive activity for spans; loop candidates from chroma autocorrelation; and lyric transcription (Whisper~\cite{radford2023whisper}).

The comparison runs across two dashboards. The \emph{single-song} view overlays each enabled detector as its own row beneath a shared timeline alongside the chosen reference (manual or AutoGuess), scored side by side by \texttt{mir\_eval} and the ambiguity scheme evaluator (Section~\ref{sec:scheme-aware}). The \emph{multi-song} \emph{Dataset Evaluation} view (bottom tab) runs any subset across the corpus and aggregates per-song scores into a ranked table. Boundary detection is the default scope; cues, spans, loops, and patterns are available under an optional flag, and any marker type also remains reachable through Custom Detectors.

\paragraph{Scheme-Aware Scoring}
\label{sec:scheme-aware}
Beyond \texttt{mir\_eval}, a scheme-aware evaluator adds matching rules tuned to multi-candidate and criticality (Section~\ref{sec:annotation-scheme}), parameterized by tolerance $\tau$ and optional weight $w_{\mathrm{o}} \in [0,1]$ (default $0.5$). \textit{Multi-candidate:} a prediction within $\tau$ of any reference candidate counts as a hit. \textit{Criticality:} each reference $g$ has $w(g)\!=\!1$ if critical, $w_{\mathrm{o}}$ if optional; recall and Mean Nearest-Boundary Distance are weighted by $w(g)$, and \emph{Critical Section Recall} reports the fraction of critical references hit within $\tau$.

\paragraph{Custom Detectors \& AutoGuess}
\label{sec:custom-detectors}

The Playground lets researchers prototype new detection ideas in a browser-based editor. A \emph{Custom Detector} is a short Python script that can emit any marker type and plays two roles at once: an algorithm candidate ranked alongside the bundled baselines, and an annotation assistant whose predictions surface in the Annotator Tool for accept\slash reject.

\label{sec:autoguess}
\emph{AutoGuess} folds a chosen subset of algorithms into one by clustering predictions within a window, keeping clusters above an agreement threshold, and surfacing each as a review card on the timeline badged with how many algorithms agreed; as a tunable baseline, sweeping window, threshold, $\tau$, and centroid method ranks consensus configurations by F1 Score.

\paragraph{System.}
\label{sec:system}
TimeCues ships as a multi-arch Docker image (amd64\slash arm64) and brings up the full system with a single \texttt{docker compose up}; the design choices behind the multi-service split, multi-arch build, detector sandbox, and shared feature cache are documented in the project repository.

\paragraph{Intended Audience \& Applications.}
TimeCues serves three audiences. \textbf{Music-analysis researchers} use Algorithm Inspect, the evaluator, and the Custom Detector hook to compare structure detectors against multi-annotator references. \textbf{Annotation teams} use the Team Dashboard, the multi-candidate scheme, and live agreement tracking to coordinate work that would otherwise sit in offline scripts. \textbf{Show engineers} (lighting, video sync, DJ tools, mashups) use the critical\slash optional and multi-candidate fields to record the timing tolerance each event needs---tight on a drop, loose on a side fill.

\enlargethispage{4.5\baselineskip}
\paragraph{Reusability \& Configuration.}
We have described only the core of the tool here, for space. More features are available and can be configured through the settings page. A user manual and guided walkthrough are bundled in-app.
Project page: \url{sapirca.github.io/timecues-studio}; source: \url{github.com/sapirca/timecues-studio}; setup and usage docs are in \codepath{./INSTALL.md} and \codepath{./docs/USER_GUIDE.md}.

\balance
\clearpage
\bibliographystyle{ACM-Reference-Format}
\bibliography{refs}


\begin{thebibliography}{57}


\ifx \showCODEN    \undefined \def \showCODEN     #1{\unskip}     \fi
\ifx \showISBNx    \undefined \def \showISBNx     #1{\unskip}     \fi
\ifx \showISBNxiii \undefined \def \showISBNxiii  #1{\unskip}     \fi
\ifx \showISSN     \undefined \def \showISSN      #1{\unskip}     \fi
\ifx \showLCCN     \undefined \def \showLCCN      #1{\unskip}     \fi
\ifx \shownote     \undefined \def \shownote      #1{#1}          \fi
\ifx \showarticletitle \undefined \def \showarticletitle #1{#1}   \fi
\ifx \showURL      \undefined \def \showURL       {\relax}        \fi
\providecommand\bibfield[2]{#2}
\providecommand\bibinfo[2]{#2}
\providecommand\natexlab[1]{#1}
\providecommand\showeprint[2][]{arXiv:#2}
\makeatletter
\@ifundefined{NAT@parse@date}{}{\let\NAT@parse@date@orig\NAT@parse@date}
\@ifundefined{NAT@parse@date}{}{\def\NAT@parse@date#1#2#3#4#5#6@@{\NAT@parse@date@orig#1#2#3#4#5#6@@\def\NAT@tempyear{0000}\def\NAT@tempexlab{{?}}\ifx\NAT@year\NAT@tempyear\ifx\NAT@exlab\NAT@tempexlab\def\NAT@date{[n.\,d.]}\else\edef\NAT@date{[n.\,d.]\NAT@exlab}\fi\fi}}
\makeatother

\bibitem[{Audacity Team}(0000)]%
        {audacity}
\bibfield{author}{\bibinfo{person}{{Audacity Team}}.} \bibinfo{year}{[n.\,d.]}\natexlab{}.
\newblock \bibinfo{title}{Audacity: Free, Open Source, Cross-Platform Audio Software}.
\newblock \bibinfo{howpublished}{\url{https://www.audacityteam.org}}.
\newblock
\newblock
\shownote{Label track import\slash export format used for time-stamped annotations.}.


\bibitem[Bittner et~al\mbox{.}(2022)]%
        {bittner2022basicpitch}
\bibfield{author}{\bibinfo{person}{Rachel~M. Bittner}, \bibinfo{person}{Juan~Jos{\'e} Bosch}, \bibinfo{person}{David Rubinstein}, \bibinfo{person}{Gabriel Meseguer-Brocal}, {and} \bibinfo{person}{Sebastian Ewert}.} \bibinfo{year}{2022}\natexlab{}.
\newblock \showarticletitle{A Lightweight Instrument-Agnostic Model for Polyphonic Note Transcription and Multipitch Estimation}. In \bibinfo{booktitle}{\emph{Proceedings of the IEEE International Conference on Acoustics, Speech and Signal Processing (ICASSP)}}. \bibinfo{publisher}{IEEE}, \bibinfo{address}{Singapore}, \bibinfo{pages}{781--785}.
\newblock
\href{https://doi.org/10.1109/ICASSP43922.2022.9746549}{doi:\nolinkurl{10.1109/ICASSP43922.2022.9746549}}


\bibitem[B{\"o}ck et~al\mbox{.}(2016)]%
        {bock2016madmom}
\bibfield{author}{\bibinfo{person}{Sebastian B{\"o}ck}, \bibinfo{person}{Filip Korzeniowski}, \bibinfo{person}{Jan Schl{\"u}ter}, \bibinfo{person}{Florian Krebs}, {and} \bibinfo{person}{Gerhard Widmer}.} \bibinfo{year}{2016}\natexlab{}.
\newblock \showarticletitle{{madmom}: A New {Python} Audio and Music Signal Processing Library}. In \bibinfo{booktitle}{\emph{Proceedings of the 24th ACM International Conference on Multimedia}}. \bibinfo{publisher}{Association for Computing Machinery}, \bibinfo{address}{Amsterdam, The Netherlands}, \bibinfo{pages}{1174--1178}.
\newblock
\showISBNx{9781450336031}
\href{https://doi.org/10.1145/2964284.2973795}{doi:\nolinkurl{10.1145/2964284.2973795}}


\bibitem[Boersma and Van~Heuven(2001)]%
        {boersma2001praat}
\bibfield{author}{\bibinfo{person}{Paul Boersma} {and} \bibinfo{person}{Vincent Van~Heuven}.} \bibinfo{year}{2001}\natexlab{}.
\newblock \showarticletitle{Speak and un{S}peak with {PRAAT}}.
\newblock \bibinfo{journal}{\emph{Glot International}} \bibinfo{volume}{5}, \bibinfo{number}{9/10} (\bibinfo{year}{2001}), \bibinfo{pages}{341--347}.
\newblock


\bibitem[Cannam et~al\mbox{.}(2010)]%
        {cannam2010sonic}
\bibfield{author}{\bibinfo{person}{Chris Cannam}, \bibinfo{person}{Christian Landone}, {and} \bibinfo{person}{Mark Sandler}.} \bibinfo{year}{2010}\natexlab{}.
\newblock \showarticletitle{{Sonic Visualiser}: an open source application for viewing, analysing, and annotating music audio files}. In \bibinfo{booktitle}{\emph{Proceedings of the 18th ACM International Conference on Multimedia}} \emph{(\bibinfo{series}{MM '10})}. \bibinfo{publisher}{Association for Computing Machinery}, \bibinfo{address}{New York, NY, USA}, \bibinfo{pages}{1467--1468}.
\newblock
\showISBNx{9781605589336}
\href{https://doi.org/10.1145/1873951.1874248}{doi:\nolinkurl{10.1145/1873951.1874248}}


\bibitem[Caparrini et~al\mbox{.}(2020)]%
        {caparrini2020automatic}
\bibfield{author}{\bibinfo{person}{Antonio Caparrini}, \bibinfo{person}{Javier Arroyo}, \bibinfo{person}{Laura P{\'e}rez-Molina}, {and} \bibinfo{person}{Jaime S{\'a}nchez-Hern{\'a}ndez}.} \bibinfo{year}{2020}\natexlab{}.
\newblock \showarticletitle{Automatic subgenre classification in an electronic dance music taxonomy}.
\newblock \bibinfo{journal}{\emph{Journal of New Music Research}} \bibinfo{volume}{49}, \bibinfo{number}{3} (\bibinfo{year}{2020}), \bibinfo{pages}{269--284}.
\newblock
\href{https://doi.org/10.1080/09298215.2020.1761399}{doi:\nolinkurl{10.1080/09298215.2020.1761399}}


\bibitem[{Cockos Inc.}(0000)]%
        {reaper}
\bibfield{author}{\bibinfo{person}{{Cockos Inc.}}} \bibinfo{year}{[n.\,d.]}\natexlab{}.
\newblock \bibinfo{title}{{REAPER}: Digital Audio Workstation}.
\newblock \bibinfo{howpublished}{\url{https://www.reaper.fm}}.
\newblock
\newblock
\shownote{Project files used to round-trip markers and regions.}.


\bibitem[D{\'e}fossez et~al\mbox{.}(2019)]%
        {defossez2019demucs}
\bibfield{author}{\bibinfo{person}{Alexandre D{\'e}fossez}, \bibinfo{person}{Nicolas Usunier}, \bibinfo{person}{L{\'e}on Bottou}, {and} \bibinfo{person}{Francis Bach}.} \bibinfo{year}{2019}\natexlab{}.
\newblock \bibinfo{title}{Music Source Separation in the Waveform Domain}.
\newblock
\showeprint[arxiv]{1911.13254}~[cs.SD]
\urldef\tempurl%
\url{https://arxiv.org/abs/1911.13254}
\showURL{%
\tempurl}


\bibitem[Egozy and Clester(2022)]%
        {egozy2019concertcue}
\bibfield{author}{\bibinfo{person}{Eran Egozy} {and} \bibinfo{person}{Ian Clester}.} \bibinfo{year}{2022}\natexlab{}.
\newblock \showarticletitle{Computer-Assisted Measure Detection in a Music Score-Following Application}. In \bibinfo{booktitle}{\emph{Proceedings of the 4th International Workshop on Reading Music Systems}}. \bibinfo{address}{Online}, \bibinfo{pages}{33--36}.
\newblock


\bibitem[Foote(2000)]%
        {foote2000automatic}
\bibfield{author}{\bibinfo{person}{Jonathan Foote}.} \bibinfo{year}{2000}\natexlab{}.
\newblock \showarticletitle{Automatic Audio Segmentation Using a Measure of Audio Novelty}. In \bibinfo{booktitle}{\emph{Proceedings of the IEEE International Conference on Multimedia and Expo (ICME)}}, Vol.~\bibinfo{volume}{1}. \bibinfo{publisher}{IEEE}, \bibinfo{address}{New York, NY, USA}, \bibinfo{pages}{452--455}.
\newblock
\href{https://doi.org/10.1109/ICME.2000.869637}{doi:\nolinkurl{10.1109/ICME.2000.869637}}


\bibitem[Gang et~al\mbox{.}(2011)]%
        {GangBLRHB11}
\bibfield{author}{\bibinfo{person}{Ren Gang}, \bibinfo{person}{Gregory Bocko}, \bibinfo{person}{Justin Lundberg}, \bibinfo{person}{Stephen Roessner}, \bibinfo{person}{Dave Headlam}, {and} \bibinfo{person}{Mark~F. Bocko}.} \bibinfo{year}{2011}\natexlab{}.
\newblock \showarticletitle{A Real-Time Signal Processing Framework of Musical Expressive Feature Extraction Using Matlab}. In \bibinfo{booktitle}{\emph{Proceedings of the 12th International Society for Music Information Retrieval Conference (ISMIR)}}. \bibinfo{publisher}{International Society for Music Information Retrieval}, \bibinfo{address}{Miami, FL, USA}, \bibinfo{pages}{115--120}.
\newblock


\bibitem[Goto(2001)]%
        {goto2001audio}
\bibfield{author}{\bibinfo{person}{Masataka Goto}.} \bibinfo{year}{2001}\natexlab{}.
\newblock \showarticletitle{An audio-based real-time beat tracking system for music with or without drum-sounds}.
\newblock \bibinfo{journal}{\emph{Journal of New Music Research}} \bibinfo{volume}{30}, \bibinfo{number}{2} (\bibinfo{year}{2001}), \bibinfo{pages}{159--171}.
\newblock


\bibitem[Goto et~al\mbox{.}(2011)]%
        {goto2011songle}
\bibfield{author}{\bibinfo{person}{Masataka Goto}, \bibinfo{person}{Kazuyoshi Yoshii}, \bibinfo{person}{Hiromasa Fujihara}, \bibinfo{person}{Matthias Mauch}, {and} \bibinfo{person}{Tomoyasu Nakano}.} \bibinfo{year}{2011}\natexlab{}.
\newblock \showarticletitle{Songle: A Web Service for Active Music Listening Improved by User Contributions}. In \bibinfo{booktitle}{\emph{Proceedings of the 12th International Society for Music Information Retrieval Conference (ISMIR)}}. \bibinfo{publisher}{International Society for Music Information Retrieval}, \bibinfo{address}{Miami, FL, USA}, \bibinfo{pages}{311--316}.
\newblock


\bibitem[Grover et~al\mbox{.}(2020)]%
        {grover2020audino}
\bibfield{author}{\bibinfo{person}{Manraj~Singh Grover}, \bibinfo{person}{Pakhi Bamdev}, \bibinfo{person}{Ratin~Kumar Brala}, \bibinfo{person}{Yaman Kumar}, \bibinfo{person}{Mika Hama}, {and} \bibinfo{person}{Rajiv~Ratn Shah}.} \bibinfo{year}{2020}\natexlab{}.
\newblock \bibinfo{title}{audino: A Modern Annotation Tool for Audio and Speech}.
\newblock
\showeprint[arxiv]{2006.05236}~[cs.SD]
\urldef\tempurl%
\url{https://arxiv.org/abs/2006.05236}
\showURL{%
\tempurl}


\bibitem[Hao et~al\mbox{.}(2026)]%
        {hao2026songformerscalingmusicstructure}
\bibfield{author}{\bibinfo{person}{Chunbo Hao}, \bibinfo{person}{Ruibin Yuan}, \bibinfo{person}{Jixun Yao}, \bibinfo{person}{Qixin Deng}, \bibinfo{person}{Xinyi Bai}, \bibinfo{person}{Yanbo Wang}, \bibinfo{person}{Wei Xue}, {and} \bibinfo{person}{Lei Xie}.} \bibinfo{year}{2026}\natexlab{}.
\newblock \bibinfo{title}{SongFormer: Scaling Music Structure Analysis with Heterogeneous Supervision}.
\newblock
\showeprint[arxiv]{2510.02797}~[eess.AS]
\urldef\tempurl%
\url{https://arxiv.org/abs/2510.02797}
\showURL{%
\tempurl}


\bibitem[Heydari et~al\mbox{.}(2021)]%
        {heydari2021beatnet}
\bibfield{author}{\bibinfo{person}{Mojtaba Heydari}, \bibinfo{person}{Frank Cwitkowitz}, {and} \bibinfo{person}{Zhiyao Duan}.} \bibinfo{year}{2021}\natexlab{}.
\newblock \showarticletitle{{BeatNet}: {CRNN} and Particle Filtering for Online Joint Beat, Downbeat and Meter Tracking}. In \bibinfo{booktitle}{\emph{Proceedings of the 22nd International Society for Music Information Retrieval Conference (ISMIR)}}. \bibinfo{publisher}{International Society for Music Information Retrieval}, \bibinfo{address}{Online}, \bibinfo{pages}{270--277}.
\newblock
\urldef\tempurl%
\url{https://archives.ismir.net/ismir2021/paper/000033.pdf}
\showURL{%
\tempurl}


\bibitem[Hils(2023)]%
        {hils2023deep}
\bibfield{author}{\bibinfo{person}{Manuel Hils}.} \bibinfo{year}{2023}\natexlab{}.
\newblock \emph{\bibinfo{title}{Deep Learning for Sound-to-Light Automation in Stage Lighting Applications}}.
\newblock \bibinfo{thesistype}{Master's\ thesis}. \bibinfo{school}{Technical University of Munich (TUM)}, \bibinfo{address}{Munich, Germany}.
\newblock
\urldef\tempurl%
\url{https://collab.dvb.bayern/spaces/TUMldv/pages/191204145/Deep+Learning+for+Sound-to-Light+Automation+in+Stage+Lighting+Applications}
\showURL{%
\tempurl}


\bibitem[Hockman et~al\mbox{.}(2012)]%
        {hockman2012jungle}
\bibfield{author}{\bibinfo{person}{Jason~A. Hockman}, \bibinfo{person}{Matthew E.~P. Davies}, {and} \bibinfo{person}{Ichiro Fujinaga}.} \bibinfo{year}{2012}\natexlab{}.
\newblock \showarticletitle{One in the Jungle: Downbeat Detection in Hardcore, Jungle, and Drum and Bass}. In \bibinfo{booktitle}{\emph{Proceedings of the 13th International Society for Music Information Retrieval Conference (ISMIR)}}. \bibinfo{publisher}{{FEUP} Edi{\c{c}}{\~{o}}es}, \bibinfo{address}{Porto, Portugal}, \bibinfo{pages}{169--174}.
\newblock


\bibitem[Humphrey et~al\mbox{.}(2014)]%
        {humphrey2014jams}
\bibfield{author}{\bibinfo{person}{Eric~J. Humphrey}, \bibinfo{person}{Justin Salamon}, \bibinfo{person}{Oriol Nieto}, \bibinfo{person}{Jon Forsyth}, \bibinfo{person}{Rachel~M. Bittner}, {and} \bibinfo{person}{Juan~Pablo Bello}.} \bibinfo{year}{2014}\natexlab{}.
\newblock \showarticletitle{{JAMS}: A {JSON} Annotated Music Specification for Reproducible {MIR} Research}. In \bibinfo{booktitle}{\emph{Proceedings of the 15th International Society for Music Information Retrieval Conference (ISMIR)}}. \bibinfo{publisher}{International Society for Music Information Retrieval}, \bibinfo{address}{Taipei, Taiwan}, \bibinfo{pages}{591--596}.
\newblock
\urldef\tempurl%
\url{https://archives.ismir.net/ismir2014/paper/000355.pdf}
\showURL{%
\tempurl}


\bibitem[Kato et~al\mbox{.}(2018)]%
        {kato2018songlesync}
\bibfield{author}{\bibinfo{person}{Jun Kato}, \bibinfo{person}{Masa Ogata}, \bibinfo{person}{Takahiro Inoue}, {and} \bibinfo{person}{Masataka Goto}.} \bibinfo{year}{2018}\natexlab{}.
\newblock \showarticletitle{Songle Sync: A Large-Scale Web-based Platform for Controlling Various Devices in Synchronization with Music}. In \bibinfo{booktitle}{\emph{Proceedings of the 26th ACM International Conference on Multimedia}}. \bibinfo{publisher}{Association for Computing Machinery}, \bibinfo{address}{Seoul, Republic of Korea}, \bibinfo{pages}{1697--1705}.
\newblock
\showISBNx{9781450356657}
\href{https://doi.org/10.1145/3240508.3240619}{doi:\nolinkurl{10.1145/3240508.3240619}}


\bibitem[Kim and Pardo(2018)]%
        {kim2018hitl}
\bibfield{author}{\bibinfo{person}{Bongjun Kim} {and} \bibinfo{person}{Bryan Pardo}.} \bibinfo{year}{2018}\natexlab{}.
\newblock \showarticletitle{A Human-in-the-Loop System for Sound Event Detection and Annotation}.
\newblock \bibinfo{journal}{\emph{ACM Transactions on Interactive Intelligent Systems}} \bibinfo{volume}{8}, \bibinfo{number}{2} (\bibinfo{year}{2018}), \bibinfo{pages}{13:1--13:23}.
\newblock
\showISSN{2160-6455}
\href{https://doi.org/10.1145/3214366}{doi:\nolinkurl{10.1145/3214366}}


\bibitem[Kim and Nam(2023)]%
        {kim2023allin1}
\bibfield{author}{\bibinfo{person}{Taejun Kim} {and} \bibinfo{person}{Juhan Nam}.} \bibinfo{year}{2023}\natexlab{}.
\newblock \showarticletitle{All-in-One Metrical and Functional Structure Analysis with Neighborhood Attentions on Demixed Audio}. In \bibinfo{booktitle}{\emph{Proceedings of the IEEE Workshop on Applications of Signal Processing to Audio and Acoustics (WASPAA)}}. \bibinfo{publisher}{IEEE}, \bibinfo{address}{New Paltz, NY, USA}, \bibinfo{pages}{1--5}.
\newblock
\href{https://doi.org/10.1109/WASPAA58266.2023.10248148}{doi:\nolinkurl{10.1109/WASPAA58266.2023.10248148}}


\bibitem[Kong et~al\mbox{.}(2020)]%
        {kong2020panns}
\bibfield{author}{\bibinfo{person}{Qiuqiang Kong}, \bibinfo{person}{Yin Cao}, \bibinfo{person}{Turab Iqbal}, \bibinfo{person}{Yuxuan Wang}, \bibinfo{person}{Wenwu Wang}, {and} \bibinfo{person}{Mark~D. Plumbley}.} \bibinfo{year}{2020}\natexlab{}.
\newblock \showarticletitle{{PANNs}: Large-Scale Pretrained Audio Neural Networks for Audio Pattern Recognition}.
\newblock \bibinfo{journal}{\emph{IEEE/ACM Transactions on Audio, Speech, and Language Processing}}  \bibinfo{volume}{28} (\bibinfo{year}{2020}), \bibinfo{pages}{2880--2894}.
\newblock
\href{https://doi.org/10.1109/TASLP.2020.3030497}{doi:\nolinkurl{10.1109/TASLP.2020.3030497}}


\bibitem[Krumhansl(1990)]%
        {krumhansl1990cognitive}
\bibfield{author}{\bibinfo{person}{Carol~L. Krumhansl}.} \bibinfo{year}{1990}\natexlab{}.
\newblock \bibinfo{booktitle}{\emph{Cognitive Foundations of Musical Pitch}}.
\newblock \bibinfo{publisher}{Oxford University Press}, \bibinfo{address}{New York, NY, USA}.
\newblock
\showISBNx{9780195148367}
\href{https://doi.org/10.1093/acprof:oso/9780195148367.001.0001}{doi:\nolinkurl{10.1093/acprof:oso/9780195148367.001.0001}}


\bibitem[Kum and Nam(2019)]%
        {kum2019jdc}
\bibfield{author}{\bibinfo{person}{Sangeun Kum} {and} \bibinfo{person}{Juhan Nam}.} \bibinfo{year}{2019}\natexlab{}.
\newblock \showarticletitle{Joint Detection and Classification of Singing Voice Melody Using Convolutional Recurrent Neural Networks}.
\newblock \bibinfo{journal}{\emph{Applied Sciences}} \bibinfo{volume}{9}, \bibinfo{number}{7} (\bibinfo{year}{2019}), \bibinfo{pages}{1324}.
\newblock
\showISSN{2076-3417}
\href{https://doi.org/10.3390/app9071324}{doi:\nolinkurl{10.3390/app9071324}}


\bibitem[Levy et~al\mbox{.}(2019)]%
        {levy2019gecko}
\bibfield{author}{\bibinfo{person}{Golan Levy}, \bibinfo{person}{Raquel Sitman}, \bibinfo{person}{Ido Amir}, \bibinfo{person}{Eduard Golshtein}, \bibinfo{person}{Ran Mochary}, \bibinfo{person}{Eilon Reshef}, \bibinfo{person}{Roi Reichart}, {and} \bibinfo{person}{Omri Allouche}.} \bibinfo{year}{2019}\natexlab{}.
\newblock \showarticletitle{{GECKO} --- A Tool for Effective Annotation of Human Conversations}. In \bibinfo{booktitle}{\emph{Proceedings of the 20th Annual Conference of the International Speech Communication Association (Interspeech)}}. \bibinfo{publisher}{ISCA}, \bibinfo{address}{Graz, Austria}, \bibinfo{pages}{3677--3678}.
\newblock
\showISSN{2958-1796}


\bibitem[Marmoret et~al\mbox{.}(2020)]%
        {marmoret2020ntd}
\bibfield{author}{\bibinfo{person}{Axel Marmoret}, \bibinfo{person}{J{\'e}r{\'e}my~E. Cohen}, \bibinfo{person}{Fr{\'e}d{\'e}ric Bimbot}, {and} \bibinfo{person}{Nancy Bertin}.} \bibinfo{year}{2020}\natexlab{}.
\newblock \showarticletitle{Uncovering audio patterns in music with Nonnegative Tucker Decomposition for structural segmentation}. In \bibinfo{booktitle}{\emph{Proceedings of the 21st International Society for Music Information Retrieval Conference (ISMIR)}}. \bibinfo{publisher}{International Society for Music Information Retrieval}, \bibinfo{address}{Montr{\'e}al, Canada}, \bibinfo{pages}{788--794}.
\newblock
\urldef\tempurl%
\url{https://archives.ismir.net/ismir2020/paper/000078.pdf}
\showURL{%
\tempurl}


\bibitem[Mauch et~al\mbox{.}(2015)]%
        {mauch2015tony}
\bibfield{author}{\bibinfo{person}{Matthias Mauch}, \bibinfo{person}{Chris Cannam}, \bibinfo{person}{Rachel~M. Bittner}, \bibinfo{person}{George Fazekas}, \bibinfo{person}{Justin Salamon}, \bibinfo{person}{Jiajie Dai}, \bibinfo{person}{Juan~Pablo Bello}, {and} \bibinfo{person}{Simon Dixon}.} \bibinfo{year}{2015}\natexlab{}.
\newblock \showarticletitle{Computer-Aided Melody Note Transcription Using the {Tony} Software: Accuracy and Efficiency}. In \bibinfo{booktitle}{\emph{Proceedings of the First International Conference on Technologies for Music Notation and Representation (TENOR)}}. \bibinfo{publisher}{Institut de Recherche en Musicologie}, \bibinfo{address}{Paris, France}, \bibinfo{pages}{23--30}.
\newblock
\showISBNx{978-2-9552905-0-7}


\bibitem[McFee and Ellis(2014a)]%
        {mcfee2014olda}
\bibfield{author}{\bibinfo{person}{Brian McFee} {and} \bibinfo{person}{Daniel P.~W. Ellis}.} \bibinfo{year}{2014}\natexlab{a}.
\newblock \showarticletitle{Analyzing Song Structure with Spectral Clustering}. In \bibinfo{booktitle}{\emph{Proceedings of the 15th International Society for Music Information Retrieval Conference (ISMIR)}}. \bibinfo{publisher}{International Society for Music Information Retrieval}, \bibinfo{address}{Taipei, Taiwan}, \bibinfo{pages}{405--410}.
\newblock


\bibitem[McFee and Ellis(2014b)]%
        {mcfee2014learning}
\bibfield{author}{\bibinfo{person}{Brian McFee} {and} \bibinfo{person}{Daniel P.~W. Ellis}.} \bibinfo{year}{2014}\natexlab{b}.
\newblock \showarticletitle{Learning to Segment Songs with Ordinal Linear Discriminant Analysis}. In \bibinfo{booktitle}{\emph{Proceedings of the IEEE International Conference on Acoustics, Speech and Signal Processing (ICASSP)}}. \bibinfo{publisher}{IEEE}, \bibinfo{address}{Florence, Italy}, \bibinfo{pages}{5197--5201}.
\newblock
\href{https://doi.org/10.1109/ICASSP.2014.6854594}{doi:\nolinkurl{10.1109/ICASSP.2014.6854594}}


\bibitem[McFee et~al\mbox{.}(2017)]%
        {mcfee2017hierarchy}
\bibfield{author}{\bibinfo{person}{Brian McFee}, \bibinfo{person}{Oriol Nieto}, \bibinfo{person}{Morwaread~M. Farbood}, {and} \bibinfo{person}{Juan~Pablo Bello}.} \bibinfo{year}{2017}\natexlab{}.
\newblock \showarticletitle{Evaluating Hierarchical Structure in Music Annotations}.
\newblock \bibinfo{journal}{\emph{Frontiers in Psychology}}  \bibinfo{volume}{8} (\bibinfo{year}{2017}), \bibinfo{pages}{1337}.
\newblock
\href{https://doi.org/10.3389/fpsyg.2017.01337}{doi:\nolinkurl{10.3389/fpsyg.2017.01337}}


\bibitem[McFee et~al\mbox{.}(2015)]%
        {mcfee2015librosa}
\bibfield{author}{\bibinfo{person}{Brian McFee}, \bibinfo{person}{Colin Raffel}, \bibinfo{person}{Dawen Liang}, \bibinfo{person}{Daniel P.~W. Ellis}, \bibinfo{person}{Matt McVicar}, \bibinfo{person}{Eric Battenberg}, {and} \bibinfo{person}{Oriol Nieto}.} \bibinfo{year}{2015}\natexlab{}.
\newblock \showarticletitle{{librosa}: Audio and Music Signal Analysis in {Python}}. In \bibinfo{booktitle}{\emph{Proceedings of the 14th Python in Science Conference (SciPy)}}. \bibinfo{publisher}{scipy.org}, \bibinfo{address}{Austin, TX, USA}, \bibinfo{pages}{18--24}.
\newblock
\href{https://doi.org/10.25080/Majora-7b98e3ed-003}{doi:\nolinkurl{10.25080/Majora-7b98e3ed-003}}


\bibitem[Mel{\'e}ndez-Catal{\'a}n et~al\mbox{.}(2017)]%
        {melendez2017bat}
\bibfield{author}{\bibinfo{person}{Blai Mel{\'e}ndez-Catal{\'a}n}, \bibinfo{person}{Emilio Molina}, {and} \bibinfo{person}{Emilia G{\'o}mez~Guti{\'e}rrez}.} \bibinfo{year}{2017}\natexlab{}.
\newblock \showarticletitle{{BAT}: An open-source, web-based audio events annotation tool}. In \bibinfo{booktitle}{\emph{Proceedings of the 3rd International Web Audio Conference (WAC)}}. \bibinfo{address}{London, United Kingdom}.
\newblock


\bibitem[{MIDI Manufacturers Association}(1996)]%
        {midi_spec}
\bibfield{author}{\bibinfo{person}{{MIDI Manufacturers Association}}.} \bibinfo{year}{1996}\natexlab{}.
\newblock \bibinfo{title}{The Complete {MIDI 1.0} Detailed Specification}.
\newblock \bibinfo{howpublished}{\url{https://www.midi.org}}.
\newblock


\bibitem[{Mixxx Development Team}(0000)]%
        {mixxx_waveforms}
\bibfield{author}{\bibinfo{person}{{Mixxx Development Team}}.} \bibinfo{year}{[n.\,d.]}\natexlab{}.
\newblock \bibinfo{title}{Mixxx: Multi-Band Frequency-Coloured Waveform Renderers}.
\newblock \bibinfo{howpublished}{\url{https://mixxx.org}}.
\newblock


\bibitem[M{\"u}ller(2015)]%
        {muller2015fundamentals}
\bibfield{author}{\bibinfo{person}{Meinard M{\"u}ller}.} \bibinfo{year}{2015}\natexlab{}.
\newblock \bibinfo{booktitle}{\emph{Fundamentals of Music Processing: Audio, Analysis, Algorithms, Applications} (\bibinfo{edition}{1} ed.)}.
\newblock \bibinfo{publisher}{Springer Cham}, \bibinfo{address}{Cham, Switzerland}.
\newblock
\showISBNx{978-3-319-21945-5}
\href{https://doi.org/10.1007/978-3-319-21945-5}{doi:\nolinkurl{10.1007/978-3-319-21945-5}}


\bibitem[Nakano and Goto(2024)]%
        {nakano2024irt}
\bibfield{author}{\bibinfo{person}{Tomoyasu Nakano} {and} \bibinfo{person}{Masataka Goto}.} \bibinfo{year}{2024}\natexlab{}.
\newblock \showarticletitle{Using Item Response Theory to Aggregate Music Annotation Results of Multiple Annotators}. In \bibinfo{booktitle}{\emph{Proceedings of the 25th International Society for Music Information Retrieval Conference (ISMIR)}}. \bibinfo{publisher}{International Society for Music Information Retrieval}, \bibinfo{address}{San Francisco, USA}, \bibinfo{pages}{1076--1084}.
\newblock


\bibitem[{Native Instruments}(0000)]%
        {ni_traktor_waveform}
\bibfield{author}{\bibinfo{person}{{Native Instruments}}.} \bibinfo{year}{[n.\,d.]}\natexlab{}.
\newblock \bibinfo{title}{Traktor {Pro}: Frequency-Coloured Waveform Display}.
\newblock \bibinfo{howpublished}{\url{https://www.native-instruments.com/traktor}}.
\newblock


\bibitem[Nieto and Bello(2016)]%
        {nieto2016msaf}
\bibfield{author}{\bibinfo{person}{Oriol Nieto} {and} \bibinfo{person}{Juan~Pablo Bello}.} \bibinfo{year}{2016}\natexlab{}.
\newblock \showarticletitle{Systematic Exploration of Computational Music Structure Research}. In \bibinfo{booktitle}{\emph{Proceedings of the 17th International Society for Music Information Retrieval Conference (ISMIR)}}. \bibinfo{publisher}{International Society for Music Information Retrieval}, \bibinfo{address}{New York, NY, USA}, \bibinfo{pages}{547--553}.
\newblock
\urldef\tempurl%
\url{https://archives.ismir.net/ismir2016/paper/000043.pdf}
\showURL{%
\tempurl}


\bibitem[Nieto and Jehan(2013)]%
        {nieto2013cnmf}
\bibfield{author}{\bibinfo{person}{Oriol Nieto} {and} \bibinfo{person}{Tristan Jehan}.} \bibinfo{year}{2013}\natexlab{}.
\newblock \showarticletitle{Convex Non-Negative Matrix Factorization for Automatic Music Structure Identification}. In \bibinfo{booktitle}{\emph{Proceedings of the 38th IEEE International Conference on Acoustics, Speech and Signal Processing (ICASSP)}}. \bibinfo{publisher}{IEEE}, \bibinfo{address}{Vancouver, Canada}, \bibinfo{pages}{236--240}.
\newblock
\href{https://doi.org/10.1109/ICASSP.2013.6637644}{doi:\nolinkurl{10.1109/ICASSP.2013.6637644}}


\bibitem[Nieto et~al\mbox{.}(2020)]%
        {nieto2020audio}
\bibfield{author}{\bibinfo{person}{Oriol Nieto}, \bibinfo{person}{Gautham~J. Mysore}, \bibinfo{person}{Cheng-i Wang}, \bibinfo{person}{Jordan B.~L. Smith}, \bibinfo{person}{Jan Schl{\"u}ter}, \bibinfo{person}{Thomas Grill}, {and} \bibinfo{person}{Brian McFee}.} \bibinfo{year}{2020}\natexlab{}.
\newblock \showarticletitle{Audio-Based Music Structure Analysis: Current Trends, Open Challenges, and Applications}.
\newblock \bibinfo{journal}{\emph{Transactions of the International Society for Music Information Retrieval}} \bibinfo{volume}{3}, \bibinfo{number}{1} (\bibinfo{year}{2020}), \bibinfo{pages}{246--263}.
\newblock
\href{https://doi.org/10.5334/tismir.54}{doi:\nolinkurl{10.5334/tismir.54}}


\bibitem[{Pioneer DJ Corporation}(0000)]%
        {pioneer_rekordbox_waveform}
\bibfield{author}{\bibinfo{person}{{Pioneer DJ Corporation}}.} \bibinfo{year}{[n.\,d.]}\natexlab{}.
\newblock \bibinfo{title}{Rekordbox: {3-Band} Frequency-Coloured Waveform Display}.
\newblock \bibinfo{howpublished}{\url{https://rekordbox.com}}.
\newblock


\bibitem[Prokopiou et~al\mbox{.}(2026)]%
        {prokopiou2026labelbuddy}
\bibfield{author}{\bibinfo{person}{Ioannis Prokopiou}, \bibinfo{person}{Ioannis Sina}, \bibinfo{person}{Agisilaos Kounelis}, \bibinfo{person}{Pantelis Vikatos}, {and} \bibinfo{person}{Themos Stafylakis}.} \bibinfo{year}{2026}\natexlab{}.
\newblock \showarticletitle{{L}abel{B}uddy: An Open Source Music and Audio Language Annotation Tagging Tool Using {AI} Assistance}. In \bibinfo{booktitle}{\emph{Proceedings of the 4th Workshop on {NLP} for Music and Audio ({NLP}4{M}us{A} 2026)}}. \bibinfo{publisher}{Association for Computational Linguistics}, \bibinfo{address}{Rabat, Morocco}, \bibinfo{pages}{7--12}.
\newblock
\showISBNx{979-8-89176-369-2}
\href{https://doi.org/10.18653/v1/2026.nlp4musa-1.2}{doi:\nolinkurl{10.18653/v1/2026.nlp4musa-1.2}}


\bibitem[Radford et~al\mbox{.}(2023)]%
        {radford2023whisper}
\bibfield{author}{\bibinfo{person}{Alec Radford}, \bibinfo{person}{Jong~Wook Kim}, \bibinfo{person}{Tao Xu}, \bibinfo{person}{Greg Brockman}, \bibinfo{person}{Christine McLeavey}, {and} \bibinfo{person}{Ilya Sutskever}.} \bibinfo{year}{2023}\natexlab{}.
\newblock \showarticletitle{Robust Speech Recognition via Large-Scale Weak Supervision}. In \bibinfo{booktitle}{\emph{Proceedings of the 40th International Conference on Machine Learning}} \emph{(\bibinfo{series}{Proceedings of Machine Learning Research}, Vol.~\bibinfo{volume}{202})}. \bibinfo{publisher}{PMLR}, \bibinfo{address}{Honolulu, HI, USA}, \bibinfo{pages}{28492--28518}.
\newblock
\urldef\tempurl%
\url{https://proceedings.mlr.press/v202/radford23a.html}
\showURL{%
\tempurl}


\bibitem[Raffel et~al\mbox{.}(2014)]%
        {raffel2014mireval}
\bibfield{author}{\bibinfo{person}{Colin Raffel}, \bibinfo{person}{Brian McFee}, \bibinfo{person}{Eric~J. Humphrey}, \bibinfo{person}{Justin Salamon}, \bibinfo{person}{Oriol Nieto}, \bibinfo{person}{Dawen Liang}, {and} \bibinfo{person}{Daniel P.~W. Ellis}.} \bibinfo{year}{2014}\natexlab{}.
\newblock \showarticletitle{{mir\_eval}: A Transparent Implementation of Common {MIR} Metrics}. In \bibinfo{booktitle}{\emph{Proceedings of the 15th International Society for Music Information Retrieval Conference (ISMIR)}}. \bibinfo{publisher}{International Society for Music Information Retrieval}, \bibinfo{address}{Taipei, Taiwan}, \bibinfo{pages}{367--372}.
\newblock
\urldef\tempurl%
\url{https://archives.ismir.net/ismir2014/paper/000320.pdf}
\showURL{%
\tempurl}


\bibitem[Ren et~al\mbox{.}(2017)]%
        {ren2017consensus}
\bibfield{author}{\bibinfo{person}{Iris~Yuping Ren}, \bibinfo{person}{Hendrik~Vincent Koops}, \bibinfo{person}{Anja Volk}, {and} \bibinfo{person}{Wouter Swierstra}.} \bibinfo{year}{2017}\natexlab{}.
\newblock \showarticletitle{In Search of the Consensus Among Musical Pattern Discovery Algorithms}. In \bibinfo{booktitle}{\emph{Proceedings of the 18th International Society for Music Information Retrieval Conference (ISMIR)}}. \bibinfo{publisher}{International Society for Music Information Retrieval}, \bibinfo{address}{Suzhou, China}, \bibinfo{pages}{671--678}.
\newblock
\urldef\tempurl%
\url{https://archives.ismir.net/ismir2017/paper/000120.pdf}
\showURL{%
\tempurl}


\bibitem[Rocha et~al\mbox{.}(2013)]%
        {rocha2013segmentation}
\bibfield{author}{\bibinfo{person}{Bruno Rocha}, \bibinfo{person}{Niels Bogaards}, {and} \bibinfo{person}{Aline Honingh}.} \bibinfo{year}{2013}\natexlab{}.
\newblock \showarticletitle{Segmentation and Timbre Similarity in Electronic Dance Music}. In \bibinfo{booktitle}{\emph{Proceedings of the 10th Sound and Music Computing Conference (SMC)}}. \bibinfo{address}{Stockholm, Sweden}, \bibinfo{pages}{754--761}.
\newblock
\href{https://doi.org/10.5281/zenodo.850377}{doi:\nolinkurl{10.5281/zenodo.850377}}


\bibitem[S{\'a}~Pinto(2025)]%
        {pinto2025maracatu}
\bibfield{author}{\bibinfo{person}{Ant{\'o}nio S{\'a}~Pinto}.} \bibinfo{year}{2025}\natexlab{}.
\newblock \showarticletitle{Towards Human-in-the-Loop Onset Detection: A Transfer Learning Approach for {Maracatu}}. In \bibinfo{booktitle}{\emph{Proceedings of the 26th International Society for Music Information Retrieval Conference (ISMIR)}}. \bibinfo{publisher}{International Society for Music Information Retrieval}, \bibinfo{address}{Daejeon, South Korea}, \bibinfo{pages}{320--327}.
\newblock
\showeprint[arxiv]{2507.04858}~[cs.SD]
\urldef\tempurl%
\url{https://arxiv.org/abs/2507.04858}
\showURL{%
\tempurl}


\bibitem[{Silero Team}(2024)]%
        {silerovad}
\bibfield{author}{\bibinfo{person}{{Silero Team}}.} \bibinfo{year}{2024}\natexlab{}.
\newblock \bibinfo{title}{{Silero VAD}: pre-trained enterprise-grade Voice Activity Detector ({VAD}), Number Detector and Language Classifier}.
\newblock \bibinfo{howpublished}{\url{https://github.com/snakers4/silero-vad}}.
\newblock


\bibitem[Smith et~al\mbox{.}(2011)]%
        {smith2011salami}
\bibfield{author}{\bibinfo{person}{Jordan B.~L. Smith}, \bibinfo{person}{John~Ashley Burgoyne}, \bibinfo{person}{Ichiro Fujinaga}, \bibinfo{person}{David De~Roure}, {and} \bibinfo{person}{J.~Stephen Downie}.} \bibinfo{year}{2011}\natexlab{}.
\newblock \showarticletitle{Design and Creation of a Large-Scale Database of Structural Annotations}. In \bibinfo{booktitle}{\emph{Proceedings of the 12th International Society for Music Information Retrieval Conference (ISMIR)}}. \bibinfo{publisher}{International Society for Music Information Retrieval}, \bibinfo{address}{Miami, FL, USA}, \bibinfo{pages}{555--560}.
\newblock
\urldef\tempurl%
\url{https://ismir2011.ismir.net/papers/PS4-14.pdf}
\showURL{%
\tempurl}


\bibitem[Solberg and Dibben(2019)]%
        {solberg2019peak}
\bibfield{author}{\bibinfo{person}{Ragnhild~Torvanger Solberg} {and} \bibinfo{person}{Nicola Dibben}.} \bibinfo{year}{2019}\natexlab{}.
\newblock \showarticletitle{Peak Experiences with Electronic Dance Music: Subjective Experiences, Physiological Responses, and Musical Characteristics of the Break Routine}.
\newblock \bibinfo{journal}{\emph{Music Perception}} \bibinfo{volume}{36}, \bibinfo{number}{4} (\bibinfo{year}{2019}), \bibinfo{pages}{371--389}.
\newblock
\href{https://doi.org/10.1525/mp.2019.36.4.371}{doi:\nolinkurl{10.1525/mp.2019.36.4.371}}


\bibitem[Truong et~al\mbox{.}(2018)]%
        {truong2018ruptures}
\bibfield{author}{\bibinfo{person}{Charles Truong}, \bibinfo{person}{Laurent Oudre}, {and} \bibinfo{person}{Nicolas Vayatis}.} \bibinfo{year}{2018}\natexlab{}.
\newblock \bibinfo{title}{ruptures: change point detection in Python}.
\newblock
\showeprint[arxiv]{1801.00826}~[stat.CO]
\urldef\tempurl%
\url{https://arxiv.org/abs/1801.00826}
\showURL{%
\tempurl}


\bibitem[Truong et~al\mbox{.}(2020)]%
        {truong2020ruptures}
\bibfield{author}{\bibinfo{person}{Charles Truong}, \bibinfo{person}{Laurent Oudre}, {and} \bibinfo{person}{Nicolas Vayatis}.} \bibinfo{year}{2020}\natexlab{}.
\newblock \showarticletitle{Selective Review of Offline Change Point Detection Methods}.
\newblock \bibinfo{journal}{\emph{Signal Processing}}  \bibinfo{volume}{167} (\bibinfo{year}{2020}), \bibinfo{pages}{107299}.
\newblock
\href{https://doi.org/10.1016/j.sigpro.2019.107299}{doi:\nolinkurl{10.1016/j.sigpro.2019.107299}}


\bibitem[Vaz et~al\mbox{.}(2016)]%
        {vaz2016convex}
\bibfield{author}{\bibinfo{person}{Colin Vaz}, \bibinfo{person}{Asterios Toutios}, {and} \bibinfo{person}{Shrikanth Narayanan}.} \bibinfo{year}{2016}\natexlab{}.
\newblock \showarticletitle{Convex Hull Convolutive Non-negative Matrix Factorization for Uncovering Temporal Patterns in Multivariate Time-Series Data}. In \bibinfo{booktitle}{\emph{Proceedings of the Annual Conference of the International Speech Communication Association (Interspeech)}}. \bibinfo{publisher}{ISCA}, \bibinfo{address}{San Francisco, CA, USA}, \bibinfo{pages}{963--967}.
\newblock
\href{https://doi.org/10.21437/Interspeech.2016-571}{doi:\nolinkurl{10.21437/Interspeech.2016-571}}


\bibitem[Wang et~al\mbox{.}(2017)]%
        {wang2017crowdsourcing}
\bibfield{author}{\bibinfo{person}{Cheng-i Wang}, \bibinfo{person}{Gautham~J. Mysore}, {and} \bibinfo{person}{Shlomo Dubnov}.} \bibinfo{year}{2017}\natexlab{}.
\newblock \showarticletitle{Re-Visiting the Music Segmentation Problem with Crowdsourcing}. In \bibinfo{booktitle}{\emph{Proceedings of the 18th International Society for Music Information Retrieval Conference (ISMIR)}}. \bibinfo{publisher}{International Society for Music Information Retrieval}, \bibinfo{address}{Suzhou, China}, \bibinfo{pages}{738--744}.
\newblock
\urldef\tempurl%
\url{https://archives.ismir.net/ismir2017/paper/000102.pdf}
\showURL{%
\tempurl}


\bibitem[Yadati et~al\mbox{.}(2014)]%
        {yadati2014drops}
\bibfield{author}{\bibinfo{person}{Karthik Yadati}, \bibinfo{person}{Martha Larson}, \bibinfo{person}{Cynthia C.~S. Liem}, {and} \bibinfo{person}{Alan Hanjalic}.} \bibinfo{year}{2014}\natexlab{}.
\newblock \showarticletitle{Detecting Drops in Electronic Dance Music: Content based approaches to a socially significant music event}. In \bibinfo{booktitle}{\emph{Proceedings of the 15th International Society for Music Information Retrieval Conference (ISMIR)}}. \bibinfo{publisher}{International Society for Music Information Retrieval}, \bibinfo{address}{Taipei, Taiwan}, \bibinfo{pages}{143--148}.
\newblock
\urldef\tempurl%
\url{https://archives.ismir.net/ismir2014/paper/000297.pdf}
\showURL{%
\tempurl}


\bibitem[Yamamoto(2021)]%
        {yamamoto2021arsa}
\bibfield{author}{\bibinfo{person}{Kazuhiko Yamamoto}.} \bibinfo{year}{2021}\natexlab{}.
\newblock \showarticletitle{Human-in-the-Loop Adaptation for Interactive Musical Beat Tracking}. In \bibinfo{booktitle}{\emph{Proceedings of the 22nd International Society for Music Information Retrieval Conference (ISMIR)}}. \bibinfo{publisher}{International Society for Music Information Retrieval}, \bibinfo{address}{Online}, \bibinfo{pages}{794--801}.
\newblock
\urldef\tempurl%
\url{https://archives.ismir.net/ismir2021/paper/000099.pdf}
\showURL{%
\tempurl}


\end{thebibliography}

\clearpage
\appendix
\section*{Appendix}

\section{Case Study: Annotating an EDM Corpus}
\label{app:case-study}
The lead author and four collaborators used TimeCues to build a 109-track EDM corpus, spanning subgenres from Afro and Organic House to Psybass, Glitch Hop, and Melodic Techno. The annotations show a consistent structural vocabulary---a median of about ten boundaries per track, dominated by drops (39\%) and buildups (23\%), then breakdowns (11\%), intros and outros (10\% each), bridges (5\%), and silences (3\%). The criticality field was actively used, with about 20\% of boundaries flagged optional and the rest critical---the hard, beat-locked transitions that must land exactly.

EDM structure is mostly \emph{band-level}, so on the frequency-colored 3-band waveform (Figure~\ref{fig:annotator}) these events read as shapes, and annotators placed most boundaries by sight, often faster than by ear. Boundaries were often ambiguous, heard at the last bar of the outgoing section or the first of the incoming one; rather than force a single timestamp, annotators recorded every defensible position with the multi-candidate field (Section~\ref{sec:annotation-scheme}). Because each annotator's layer is stored separately yet shares the beat grid, the Team Dashboard ranks tracks by inter-annotator disagreement, so the admin can target the least-agreed tracks and consolidate each into one reviewed annotation. Agreement is the pairwise tolerance-windowed boundary F1---greedy one-to-one matching within $\tau$, with label agreement over the matched boundaries---not a categorical Kappa or Alpha, which miss continuous multi-candidate alignment.

\section{Technical Architecture and Design Choices}
\label{app:architecture}

\subsection{Model-assisted annotation}
Integrating detection models directly lets TimeCues request predictions on demand and drop them onto the canvas as editable pre-annotations, shifting the task from creation to verification---the accept\slash reject loop of Section~\ref{sec:custom-detectors} (Figure~\ref{fig:detector-review}). The verified labels accumulate into a corpus for fine-tuning, closing a human-in-the-loop cycle. This matters most when pretrained models underperform on under-represented material---a beat tracker trained on steady 4/4 pop drifts on frequent tempo changes---where a few corrected examples adapt a detector with little effort~\cite{pinto2025maracatu}.

\optfigh[\linewidth]{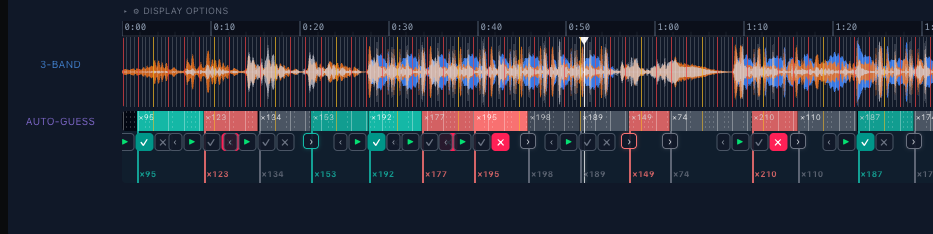}{%
  AutoGuess review band: accept\slash reject cards.}{Consensus clusters shown as accept-or-reject review cards under the shared waveform, each badged with the number of algorithms that agreed.}{fig:detector-review}

\subsection{One process per concern}
Backend capabilities are decoupled into separate services, each pinning its own dependencies. This matters because the underlying audio and deep-learning libraries pin conflicting versions of NumPy, PyTorch, and Cython, so one combined image is fragile; isolated services track their upstreams independently and restart on failure without taking the workspace down.

\subsection{Multi-arch image, single source}
The Docker image builds for \texttt{amd64} and \texttt{arm64} from one \texttt{Dockerfile} via \texttt{buildx}. Researchers develop on Apple Silicon, lab servers run Intel, the cloud runs either, and one tag covers all three. Compose profiles (\texttt{base}, \texttt{gpu}, \texttt{cpu}, \codepath{experimental-models}) add opt-in services on top of the base install, so a first-time user does not download CUDA or experimental model weights they will not use.

\subsection{Sandboxed Python detectors}
A \emph{Custom Detector} is a short Python script the researcher edits in the Playground editor (Figure~\ref{fig:custom-detectors}); hitting \emph{Run} executes it and the result lands on the canvas. Scripts run server-side in an isolated child process with memory, CPU, and wall-clock limits, so an honest mistake---a runaway loop, an accidental allocation---is contained. The same script serves both interactive single-song testing and batch evaluation across the corpus.

\optfigh{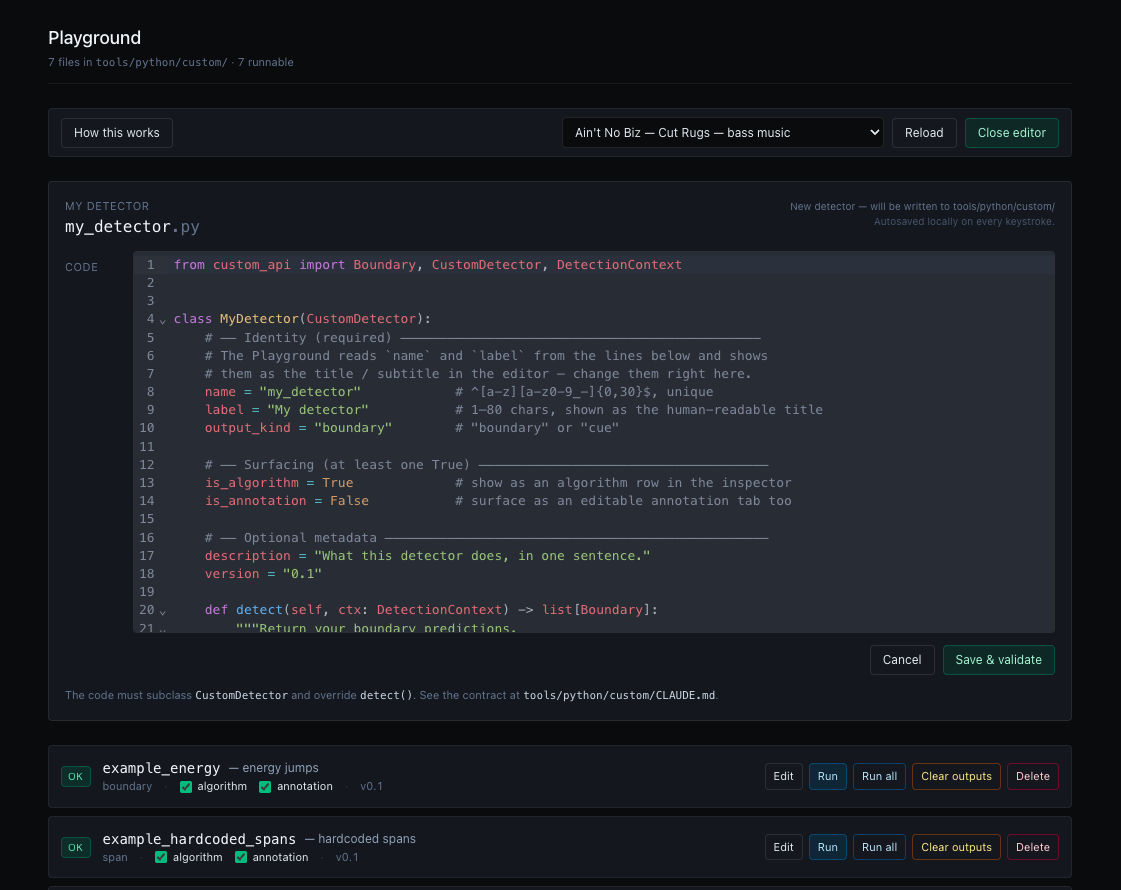}{%
  Playground: in-browser editor for Custom Detectors, which can feed the inspector, the annotator, or both.}{Playground editor above the list of bundled detectors.}{fig:custom-detectors}

\subsection{Pre-computed feature cache}
The feature server caches every expensive computation (mel-spectro\-gram, MFCC, chroma, tempogram, SSM, novelty, Demucs stems) under \codepath{(song-slug, feature-name)} keys with file-hash invalidation. Custom detectors and built-in baselines share the same cache, so a set of \texttt{ruptures} configurations and MSAF segmenters all read the same chroma frames. Adding a new configuration costs only its detection-and-scoring pass, not another round of feature extraction. Every signal offered by the \textsc{Signals} picker (Figure~\ref{fig:signal-layers}) and every Demucs stem in the source selector (Figure~\ref{fig:stem-source}) is served from this cache, so toggling a signal or switching stems is instant after the first computation.

\optfigh[0.85\linewidth]{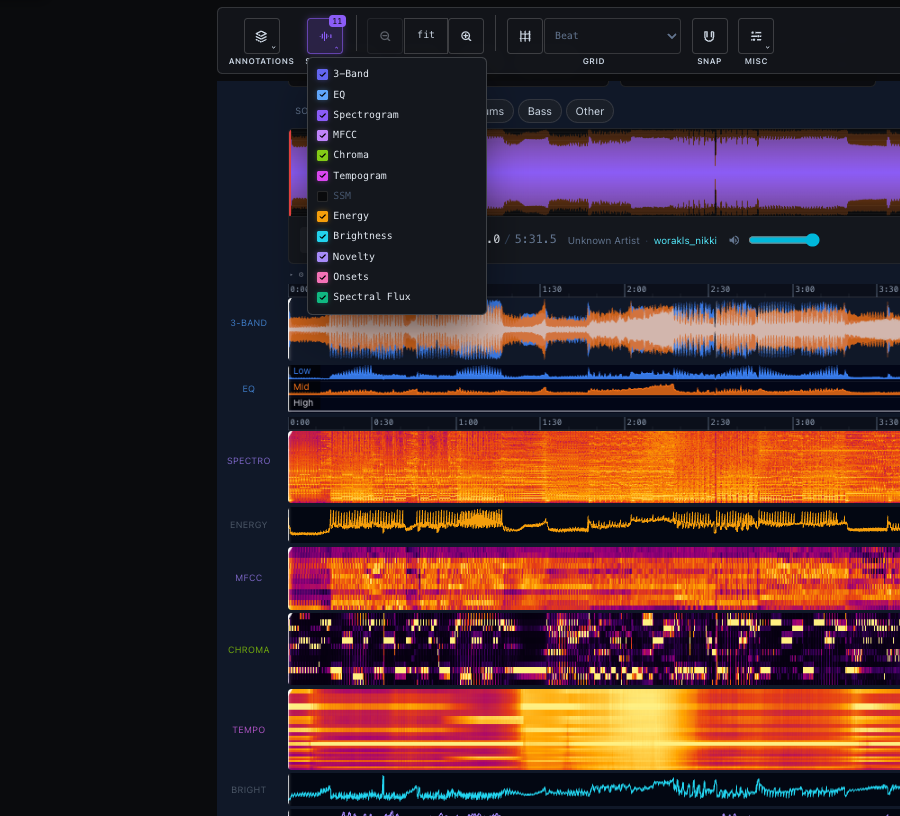}{%
  The \textsc{Signals} picker toggles analysis layers---waveform, EQ, spectrogram, MFCC, chroma, tempogram, SSM, energy, brightness, novelty, onsets, spectral flux---as a synchronized stack over the shared timeline.}{Signals dropdown open over a vertical stack of feature visualizations sharing one timeline.}{fig:signal-layers}

\optfigh{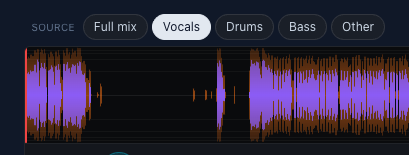}{%
  Per-stem source selector: any signal can be rendered for the full mix or a single Demucs stem (Vocals\slash Drums\slash Bass\slash Other). Vocals is selected here.}{Stem-source row with Full mix, Vocals, Drums, Bass, and Other buttons above a vocals-only waveform.}{fig:stem-source}

\subsection{Files over database}
Annotations live as JSON files on disk, one per song per layer per annotator, at
\begin{center}
\texttt{data/annotations/\textless layer\textgreater/\textless annotator\textgreater/\textless slug\textgreater.json}
\end{center}
Per-annotator subdirectories stop annotators from overwriting each other; the annotator chosen at sign-in (Figure~\ref{fig:sign-in})---via Google or a plain username\slash email---supplies the \texttt{\textless annotator\textgreater} path component. Because each category lives in its own files, a storage panel accounts for each separately and clears regenerable caches in one click (Figure~\ref{fig:storage}).

\optfigh[0.7\linewidth]{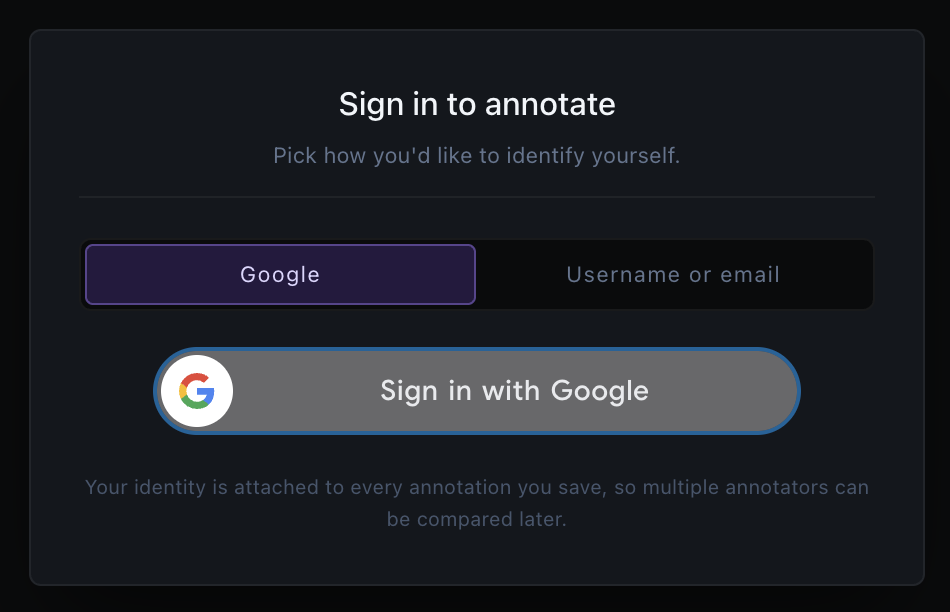}{%
  Lightweight sign-in: an annotator identifies via Google or a plain username\slash email, and that identity is attached to every saved annotation so multiple annotators' layers can be compared later.}{Sign-in modal offering Google or username/email identification before annotating.}{fig:sign-in}

\optfigh[0.7\linewidth]{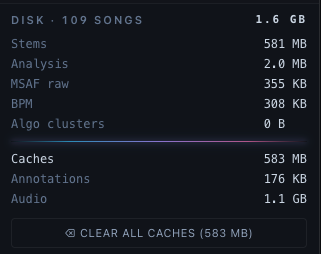}{%
  On-disk footprint for a 109-track corpus: stems, analysis, raw MSAF output, BPM, and algorithm clusters are accounted separately, and regenerable caches are cleared in one click.}{Disk-usage panel listing per-category storage for 109 songs totaling 1.6\,GB.}{fig:storage}

\subsection{Shared beat-locked canvas}
A per-song beat grid drives every view, and markers snap to it (toggleable) so a manual boundary and an algorithm prediction all line up on the same lines. The same grid serves Dataset Prep, the Annotator Tool, and Algorithm Inspect, so an alignment set once holds everywhere; grid mode and the snap toggle live on a shared canvas toolbar (Figure~\ref{fig:toolbar}).

\optfigh[\linewidth]{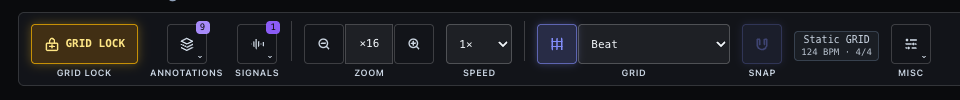}{%
  Canvas toolbar shared across all three workspaces: the \textsc{Annotations} and \textsc{Signals} layer pickers, zoom, the \textsc{Grid} mode selector, the snap toggle, and \textsc{Misc} options.}{Horizontal toolbar with Annotations, Signals, Zoom, Grid, Snap, and Misc controls.}{fig:toolbar}

\subsection{Standard formats, no lock-in}
Export covers JSON, JAMS, Audacity, Sonic Visualiser, MIDI, and REAPER; import covers JSON, JAMS, Audacity, and CSV. The scheme-aware evaluator and multi\-candidate format are specified separately from the GUI, so a team can stop using the workspace and still keep their annotations.

\subsection{Demo without server state}
A bundled demo of three CC0 tracks with pre-baked Demucs stems ships inside the image and runs client-side via \texttt{localStorage}, so the tool can be tried end-to-end without uploading audio, creating an account, or touching server-side data.

\subsection{One-command deploy}
A provided deployment pipeline builds for both architectures and ships the image.

\end{document}